**A COMPARISON OF NEW CALCULATIONS OF THE YEARLY $^{10}$Be PRODUCTION**

**IN THE EARTHS POLAR ATMOSPHERE BY COSMIC RAYS WITH YEARLY $^{10}$Be**

**MEASUREMENTS IN MULTIPLE GREENLAND ICE CORES BETWEEN 1939 AND**

**1994 – A TROUBLING LACK OF CONCORDANCE**

**PAPER #2**

**W.R. Webber[1], P.R. Higbie[2] and C.W. Webber[3]**

1. New Mexico State University, Department of Astronomy, Las Cruces, NM 88003, USA

2. New Mexico State University, Physics Department, Las Cruces, NM 88003, USA

3. Microsoft Corp., Building 88, Redmond, WA 98052, USA

# ABSTRACT

We have compared the yearly production rates of $^{10}$Be by cosmic rays in the Earths polar atmosphere over the last 50-70 years with $^{10}$Be measurements from two separate ice cores in Greenland. These ice cores provide measurements of the annual $^{10}$Be concentration and $^{10}$Be flux levels during this time. The scatter in the ice core yearly data vs. the production data is larger than the average solar 11 year production variations that are being measured. The cross correlation coefficients between the yearly $^{10}$Be production and the ice core $^{10}$Be measurements for this time period are <0.4 in all comparisons between ice core data and $^{10}$Be production, including $^{10}$Be concentrations, $^{10}$Be fluxes and in comparing the two separate ice core measurements. In fact, the cross correlation between the two ice core measurements, which should be measuring the same source, is the lowest of all, only ~0.2. These values for the correlation coefficient are all indicative of a "poor" correlation. The regression line slopes for the best fit lines between the $^{10}$Be production and the $^{10}$Be measurements used in the cross correlation analysis are all in the range 0.4-0.6. This is a particular problem for historical projections of solar activity based on ice core measurements which assume a 1:1 correspondence. We have made other tests of the correspondence between the $^{10}$Be predictions and the ice core measurements which lead to the same conclusion, namely that other influences on the ice core measurements, as large as or larger than the production changes themselves, are occurring. These influences could be climatic or instrumentally based. We suggest new ice core measurements that might help in defining more clearly what these influences are and-if possible-to correct for them.

## **Introduction**

The use of $^{10}$Be concentration measurements in polar ice cores has become an important new tool for probing the recent history of solar activity and its interaction with the Earths environment through the solar modulation of cosmic rays. The initial papers on this subject (e.g., Beer, et al., 1990, 1998) suggested that the $^{10}$Be concentration in polar ice cores, in some cases dating back a few 1,000 years, might provide a "monitor" of solar modulation activity in the same way that neutron monitors presently monitor this solar modulation level (Beer, 2000). This possibility arises because the $^{10}$Be produced in the atmosphere by cosmic rays is believed to precipitate out of the atmosphere in a uniform way in the form of rain or snow after a time period ~ 1 year. This is in contrast to $^{14}$C which is also produced by cosmic rays and is used as an indicator of solar activity in the past, but which has a much longer and more involved sequestering process.

Most of these historical studies dating back several hundred years using $^{10}$Be have considered atmospheric effects to be relatively unimportant and have therefore assumed a simple 1:1 correspondence between changes in $^{10}$Be production in the Earths atmosphere and $^{10}$Be concentration measurements in ice cores (e.g., Steinhilber, et al., 2010, and references therein). This approach is used in spite of several cautionary arguments concerning the importance of atmospheric effects (e.g., Lal, 1988; Nikitin, et al., 2005). Newer calculations using more specific atmospheric models have continued to emphasize the importance of the atmospheric contribution to the $^{10}$Be concentration measurements (e.g., Field, Schmidt and Shindell, 2009, and references therein).

In a recent paper (Webber and Higbie, 2010a) we have approached this question from a different perspective, examining the cross-correlation between yearly average $^{10}$Be concentration values in ice cores and evaluations of yearly average $^{10}$Be production by incident cosmic rays over the most recent 60-70 years. The cross-correlation coefficients we determined (~ 0.3) and the slopes of the regression lines between $^{10}$Be concentration and $^{10}$Be production (~ 0.5) both strongly suggest that other factors are modifying a simple 1:1 relationship between the $^{10}$Be production and the concentration measurements and that these other factors may be of comparable importance to the production changes themselves.

Recently, new $^{10}$Be concentration measurements have been reported from an ice core in Northern Greenland (NGRIP) (Berggren, et al., 2009). These measurements are valuable for

70    several reasons. First, they provide the opportunity to compare in detail the yearly [10]Be

71    measurements made in two cores, the original Dye-3 core and the new NGRIP core, with the

72    atmospheric [10]Be production which is known to be the same at both locations. Both ice cores are

73    in the same general region but ~1000 km apart and are on the polar plateau of latitude

74    independent cosmic ray intensities above ~60° geomagnetic latitude. Second, the new ice core

75    studies include [10]Be flux calculations as well as concentration measurements for both the NGRIP

76    core as well as the older Dye-3 core. The [10]Be flux is defined as F=CR where C is the

77    concentration and R is the snow accumulation rate determined from measurements and

78    calculations. It has been argued that the [10]Be "flux" values may provide a better representation

79    of the actual [10]Be atmospheric production since they account for the yearly snowfall differences

80    (Beer, et al., 2000). In this new paper we expand on the Webber and Higbie, 2010a paper,

81    examining the cross correlation and other features of the most recent [10]Be ice core data of

82    Berggren, et al., 2009.

## **The Data**

84    In Table 1 we show the following data. In columns B and C are the [10]Be concentration

85    and flux data from NGRIP (Berggren, et al., 2009). Column D shows the [10]Be production data

86    from Webber, Higbie and McCracken, 2007, updated to 2009. Columns E and F show revised

87    Dye-3 concentration and flux data also from Berggren, et al., 2009. Column G is a weighted

88    average of the NGRIP and Dye-3 concentration data. Column H is the yearly average sunspot

89    data through 2009 (http://www.ngdc.noaa.gov/stp/SOLAR/ftpsunspotnumber.html#american),

90    and column I is the yearly average high latitude neutron monitor rate through 2009 (with the

91    1954 rate = 100.0).

92    First we compare the yearly [10]Be production rate with the high latitude neutron monitor

93    rate. These two rates are shown in Figure 1. The cross correlation coefficient between these

94    rates is 0.995 and the slope of the normalized regression line between the two sets of data is

95    0.322 (Figure 2A). This high cross correlation is not surprising since the NM rate is used as one

96    of the bases to determine the production rate. Differences in the correlation coefficient from 1.0

97    arise from slight differences in the rigidity dependence of the solar modulation from cycle to

98    cycle.

99    The comparison of the [10]Be production rate and the sunspot number for this 70 year time

100   period is shown in Figure 2B. Here the cross correlation coefficient = 0.841 and there is more

101     scatter in the individual data points. These differences are partly caused by the differences in

102     solar modulation between the positive and negative polarity periods of the 22 year solar polarity

103     cycle. Nevertheless, the sunspot number makes an acceptable proxy for the $^{10}$Be production for

104     this restricted time period if one is interested in accuracies of 10-20%. Notice, however, that the

105     value of the maximum $^{10}$Be production at zero sunspot number is ~5.1 atoms/cm$^{2}$·sec$^{1}$. For the

106     modern epoch this is the maximum production rate of $^{10}$Be that should be expected. At the time

107     of the "Grand Maximum" of $^{10}$Be concentration at ~1700 AD, the $^{10}$Be concentration was at least

108     1.6 times higher than this (e.g., Beer, 2000; McCracken, et al., 2004). This high level of

109     concentration cannot be explained by the modern epoch sunspot data, and requires a "new"

110     regression line for production that must be a factor ~1.6 times higher for a given sunspot number

111     to describe this earlier data period.

112         Turning now to the ice core data we show in Figure 3A the scatter plot of the new NGRIP

113     yearly $^{10}$Be flux data and $^{10}$Be production in the polar atmosphere. Here the scatter is very large

114     with some extreme flux years. The cross correlation coefficient is 0.397 and the slope of the best

115     fit normalized regression line is 0.503. For a lag of ± 1 year the cross correlation coefficient

116     drops to ~0.30. For the NGRIP concentration data the cross correlation coefficient is 0.341 and

117     the slope of the normalized regression line is 0.469, slightly worse than those values for the flux,

118     and still below a cross correlation coefficient of 0.50 which is considered to be a "weak"

119     correlation and still with a slope that is much less than that for an expected 1:1 correspondence.

120         In Figure 3B the scatter plot of the Dye-3 concentration vs. $^{10}$Be production is shown.

121     Here the scatter is again very large with a significant fraction of "high" concentration years. The

122     cross correlation coefficient is now 0.306 and the slope of the best fit normalized regression line

123     is 0.573. This slope is increased as a result of a number of anomalously "high" data points (see

124     Webber and Higbie, 2010a). Again, lags of ± 1 year do not significantly alter the correlation

125     coefficient.

126         For the Dye-3 flux data in Figure 3C the correlation coefficient with $^{10}$Be production is

127     only 0.224. This is evident from the extreme scatter of the data. The slope of the best fit

128     normalized regression line is 0.437.

129         In Figure 4A we show the scatter plot of the average of the NGRIP plus 2 x Dye-3 yearly

130     concentration measurements with the $^{10}$Be production. Here the cross correlation coefficient is

131  0.414, just slightly higher than the individual NGRIP and Dye-3 concentration measurements
132  separately.  The best fit normalized regression line slope is 0.544.

133       In Figure 4B we show a scatter plot between the NGRIP and Dye-3 flux measurements
134  themselves.  This comparison should be independent of the [10]Be production calculations since
135  both NGRIP and Dye-3 would be expected to observe the same [10]Be production, whatever it is,
136  since they are both on the polar plateau for [10]Be production as noted earlier.  Here the scatter is
137  very large and the correlation coefficient = 0.256.  The regression line slope ~zero implies
138  essentially no correlation!

139       In summary, regarding the ice core measurements, we observe: 1) The scatter in the
140  individual yearly data points in the cross correlation with [10]Be production is very large in all
141  comparisons whether one considers the measured [10]Be concentrations or the deduced [10]Be
142  fluxes, 2) The corresponding cross correlation coefficients are very low.  Even the highest value
143  of 0.414 obtained for the weighted average NGRIP plus Dye-3 data is still well below the
144  criterion of 0.5 which is generally considered a weak correlation.  3) Adding the two different ice
145  core measurements or comparing the [10]Be flux determinations vs. concentration measurements of
146  [10]Be in one ice core improves the correlations only slightly.  4) The best fit normalized
147  regression line slopes between the ice core data and the [10]Be production are between ~0.4-0.6 or
148  only about ½ of the expected value of 1.0 for a direct correspondence.  5) A cross-correlation of
149  both of the yearly [10]Be concentration and flux measurements from the two sites (which should be
150  observing the same [10]Be production) give correlation coefficients less than 0.25 in both cases and
151  slopes of regression lines (which should be ~1.0) of between 0.0 and 0.3.  The level of non-
152  correlation described above requires some explanation.

153  **Discussion – The Large Scatter in the Data and the Regression Line Slopes of 0.4-0.6**
154  **Instead of 1.0**

155       The large scatter between the [10]Be ice core measurements and [10]Be production
156  calculations and the resulting lack of correlation on a year to year basis, described above, could
157  arise in several ways.  Two possibilities are; 1) Severe climatic effects on a time scale ~1-2 years
158  or less (e.g., Pedro, et al., 2006).  These effects could be very local or more general effects
159  covering large geographic scales.  2) "Instrumentally" based effects that somehow introduce a
160  large variability in what are more uniform variations.

It has been shown by Berggren, et al., 2009, and other earlier work (e.g., Beer, et al., 1998) that the historical $^{10}$Be data do show the persistence of a "11 year" variation comparable to that of sunspots (Schwabe cycle) when the $^{10}$Be data is "filtered" with a band pass of 8-16 years. This would seem to imply some level of correlation between the ice core data and the $^{10}$Be production data unless the 11 year variation in the ice core data were mainly a climatic effect. Berggren, et al., 2009, find that this 11 year $^{10}$Be variation is large during the recent time period we have studied in this paper.

Because of the extensive use of $^{10}$Be as a significant source on past solar activity, in some cases of a detailed quantitive nature (e.g., McCracken, et al., 2004), it is important to understand how the observed lack of concordance between ice-core and production data on $^{10}$Be on a yearly basis in the modern era will affect the accuracy of these quantitative historical studies. For example; Can the presently available $^{10}$Be ice-core data really prove anything about solar activity in the past; a question that has been also asked (and answered) in recent papers by Nikitin, et al., 2005, and Stozhkov, 2007.

In evaluating this ice core data it is important, not only to try to understand the source of the large scatter in the data that is observed, but also to understand why the regression line slopes between $^{10}$Be ice core data and $^{10}$Be production data are only ~ ½ of those expected for a simple 1:1 correspondence between $^{10}$Be ice core measurements and $^{10}$Be production. The value of this slope is particularly important for extrapolating the $^{10}$Be ice-core records to earlier times in a quantitative sense and it is quite different from a simple 1:1 correspondence that is now used (see Steinhilber, et al., 2010; Webber and Higbie, 2010b, and references therein).

Perhaps most significant is the result of extrapolating the observed regression lines to zero production. In all cases, this extrapolation leads to a non-zero ice core $^{10}$Be flux or concentration when the production equals zero. These non-zero $^{10}$Be values are large, ~0.5 times or greater than the values measured at the recent times of maximum modulation. These significant non-zero values are not expected, given the present "one track" $^{10}$Be models for production with subsequent distribution through the atmosphere and ultimate "precipitation" which is recorded in ice cores. If, however, the disposition of $^{10}$Be is more like that of $^{14}$C where several reservoirs are involved (e.g., atmosphere, ocean and soil) each on different time scales, then the non-zero intercept could be explained as well as the large scatter of the data between the "direct" production and ice core observations. In this case the "indirect" contribution of $^{10}$Be to

192    the ice core measurements from the other reservoirs or "channels" may be of comparable

193    importance to the "direct" production contribution.  In fact, the regression line slopes may be the

194    first indication of the importance of these "indirect" channels in the [10]Be sequestering process

195    and may alter the current viewpoint of ice core [10]Be data as providing a measure of coincident

196    [10]Be production.

197    **Discussion – Overall view of [10]Be Ice Core Data and the Solar Modulation Level**

198        In Figure 5 we show the [10]Be flux measured at NGRIP from 1600 A.D. to 1994.  The

199    data is presented as 5 year running averages.  The average maximum flux (x $10^3$) for the 5 most

200    recent periods of minimum solar (sunspot) activity is 8.55 atoms·$cm^2$·$s^{-1}$ and for the 5 most recent

201    periods of maximum activity is 6.75 atoms·$cm^2$·$s^{-1}$.  The upper bound of the shaded region

202    corresponds to a [10]Be flux which is ~1.5 times the recent average maximum [10]Be flux.  There are

203    several time periods from about 1640 to ~1892 when the [10]Be flux measured in the NGRIP ice

204    core is above this level.  At about 1695 A.D. and again about 1812 this flux level is ~2.0 times

205    the average measured at the time of the 5 most recent sunspot minima.  These are sustained time

206    intervals of high fluxes because of the 5 year running averages used.  In order for these high

207    earlier fluxes of [10]Be to be solely caused by increases in [10]Be production, assuming a 1:1

208    correspondence between ice core [10]Be and [10]Be production, requires that the [10]Be production at

209    the top of the atmosphere be at least 1.5-2.0 times the average production at the time of the 5

210    most recent sunspot minima.

211        In a recent paper, Webber and Higbie, 2010b, have shown that for the new LIS H and He

212    spectra they have derived, which are consistent with the H and He spectra measured by Voyager

213    1 beyond 110 AU, the maximum possible increase of [10]Be production above the value at the

214    Earth at the time of the five most recent maxima is, in fact, 1.5 times, or the upper limit of the

215    shaded region in Figure 5.

216        If we consider the additional fact that the slope of the observed normalized regression

217    lines between [10]Be in ice cores and [10]Be production is actually ~0.5 and not 1.0, as described

218    earlier in this paper, then we have an upper limit to the absolute maximum [10]Be flux which is

219    only ~1.25 times the recent average maximum intensity of [10]Be measured.  This value

220    corresponds to the lowest bound of the shaded region in Figure 5.  This lower bound includes

221    many other earlier time periods with [10]Be flux values that exceed those possible from [10]Be

222    production alone from the full LIS spectrum.  Indeed this implies that more than 50% the [10]Be

223  flux increase around, e.g., 1700 A.D., 1810 A.D. and 1895 A.D. is due to non-production related
224  increases!

225      Examining in more detail the most recent time period from ~1940 to 1990 when $^{10}$Be
226  production calculations, $^{10}$Be ice core measurements and sunspot observations are all available,
227  we show in Figure 6 the yearly values of these three quantities.  Here the sunspot number is
228  inverted and taken one year earlier than production to allow for the propagation of the solar wind
229  to the heliospheric termination shock.      The $^{10}$Be ice core data is taken as a 5 year running
230  average to reduce scatter in the data.

231      First we note the quite good agreement between the 11 year profiles for $^{10}$Be production
232  (blue) and sunspot number (red).  This is consistent with the relatively high cross correlation
233  coefficient ~0.85 as described earlier.  On the other hand the $^{10}$Be ice core 11 year profiles are
234  very jittery, even though 5 year running averages are used.  The amplitude of each 11 year
235  profile is quite different in the $^{10}$Be ice core data and does not correspond well to the amplitude
236  (or the more regular shape) of the corresponding 11 year profile in production or in sunspots.  In
237  particular the maximum level of $^{10}$Be flux varies from ~0.7 to 1.0 for the 5 most recent 11 year
238  cycles whereas the production itself varies by ± 5%.  And finally, SS #18 starting in ~1943
239  shows a well defined maximum (minimum) in both production and sunspot number, but
240  essentially this cycle does not exist in the NGRIP $^{10}$Be flux data (it is weakly present in the
241  NGRIP concentration data and the Dye 3 flux and concentration data).

242      All of the features described above suggest that the simple sunspot number is a much
243  more robust indicator of solar activity and solar modulation during the modern time period than
244  the current $^{10}$Be ice core measurements.

245      Turning again to the longer term studies of solar modulation, in particular the 11 year
246  (Schwabe) cycle and its persistence over time, we refer again to Figure 5.  Using the sunspot
247  number as a reference, each cycle is well defined (data before 1700 A.D. is not used because
248  there are contradictory estimates of the sunspot numbers).  There are periods of lower amplitude
249  e.g., 1790-1830 A.D. and again before ~1720 A.D. which are referred to as the Dalton and
250  Maunder minima respectively.  Note that the sunspot number does not go zero throughout these
251  entire time periods, but the amplitude of the 11 year cycle is smaller with the peak sunspot
252  numbers being ~50 instead of the normal values, 100-150.  The distribution of the lengths of
253  each sunspot cycle is sharply peaked with 9 cycles having a length of 10 years and 8 having a

254　length of 11 years (out of a total of 28 cycles).  The shortest cycle is 9 years and the longest is 14

255　years.　The cycles longer than 12 years appear to be associated with the various "Grand

256　Minimum" periods.

257　　　Regarding the variability of the amplitude and length of the 11 year cycle obtained from

258　the [10]Be ice core flux values, it is difficult to determine a pattern.  This is because some cycles

259　are either missing or absent in the [10]Be ice core data as compared with the sunspot 11 year

260　variations.  Some examples of these unusual cycles are noted by the vertical arrows shown in

261　Figure 5.  At about 1860 A.D. there was a peak in [10]Be flux associated with a peak in sunspot

262　number, the opposite to what would be expected for normal modulation conditions.  These same

263　is true at about 1829, and again at about 1884 and 1894.  In effect the [10]Be flux cycle is almost

264　exactly out of phase with the sunspot cycle in these 4 cycles.

265　　　In 1922 the opposite was observed, a sunspot minimum, but with the absence of a

266　corresponding peak in the [10]Be flux.  This is also the case in 1943 as has been noted earlier.  Of

267　the 28 sunspot minima observed since 1700 A.D., only 11 had [10]Be flux maxima at the same

268　time as the sunspot minima (± 1 year).

269　## What Can Be Extracted from the [10]Be Ice Core Data?

270　　　This question must inevitably arise because of the extreme scatter and low correlation

271　coefficients between the [10]Be production predictions and the [10]Be ice core yearly data for the

272　modern era.  This [10]Be uncertainty extends to earlier times as well as evidenced by the [10]Be

273　cyclic data in Figure 5 as described above which appears to be jittery (noisy?) in both time and

274　amplitude, particularly with respect to the many cyclic peaks that should relate to a 11 year

275　variation but correspond poorly with the much more regular 11 year sunspot cycle variation.

276　　　One does notice a certain "regularity" in the [10]Be ice core variations, however.  In Figure

277　5 we show as blue lines, the envelope of the "minima" in the [10]Be cyclic intensity variations.

278　This "envelope" has long term periodic maxima occurring at ~1685, 1815 and 1895 A.D.  This

279　type of variation could possibly be related to long term [10]Be production changes.  Indeed 22 year

280　averages (filters) of the [10]Be concentration, which smooth out the shorter term cyclic variations,

281　show broad maxima at approximately these times (McCracken, et al., 2004).  These times also,

282　more or less, coincide with the times of maxima of [14]C concentration in tree rings (Bard, et al.,

283　1997).

284    The $^{14}$C concentration appears to have a much longer and more complex time history

285 between production in the atmosphere (which is almost identical to that of $^{10}$Be) and its

286 appearance as $^{14}$C in tree rings. Nikitin, et al., 2005, have determined a cross correlation

287 coefficient of 0.49, with a lag ~6 years between the 22 year average $^{10}$Be data in Antarctica, and

288 the $^{14}$C tree ring data.

289    The $^{10}$Be data as shown in Figure 5 needs a considerably better level of understanding,

290 particularly with regard to the short term (~11 year) variability, before it should be used to

291 ascertain the absolute magnitude of possible longer and shorter term changes in $^{10}$Be production,

292 however.

293 **Summary and Conclusions**

294    In summary we observe that the scatter in the individual yearly ice core data points in the

295 cross correlation with $^{10}$Be production for the last 50-70 years is very large in all comparisons;

296 larger, in fact, than the 11 year solar modulation effect that is being determined. The

297 corresponding correlation coefficients ~0.3-0.4 are very low. These correlation coefficients are

298 well below the value ~0.5 which is generally considered a weak correlation.

299    The best fit regression line slopes between the ice core data and the $^{10}$Be production for

300 this same time period are between 0.4-0.6 or only ~ ½ of the expected value of 1.0 for a direct

301 correspondence.

302    Perhaps most troubling of all is the cross correlation of the yearly $^{10}$Be concentration and

303 flux measurements themselves from two sites on the polar plateau which should be observing the

304 same $^{10}$Be production. These cross correlation coefficients are the lowest of all, less than 0.25

305 for both concentration and flux measurements and the slopes of the regression lines are less than

306 0.3.

307    These low cross correlation coefficients and the regression line slopes ~0.5 or less, as

308 well as the fact that for zero production there is a "large" non-zero intercept for the $^{10}$Be flux or

309 concentration levels in ice cores may be indicative of the fact that current models relating $^{10}$Be

310 production and $^{10}$Be in ice cores may be too simplistic. Other reservoirs for $^{10}$Be in addition to

311 the current direct atmospheric sequestering process may be involved, similar to $^{14}$C.

312    From the point of view of the cyclic variations of the $^{10}$Be data, we find that in the 5 most

313 recent 11 year cycles where production data is available for $^{10}$Be, sunspot number data is

314 available, and yearly $^{10}$Be concentration and flux data are also available from two sites in

315    Greenland, the sunspot number provides the best determination of the $^{10}$Be production and the
316    related solar activity. The $^{10}$Be ice core data provides a "poor" basis for determining the cyclic
317    solar activity/$^{10}$Be production level and, in fact, the NGRIP data completely misses solar cycle
318    #18.

319    Extending the above comparison in the modern age to earlier times after ~1700 A.D.
320    when sunspot records were available, it is clear that the $^{10}$Be cyclic variations provide a very
321    poor representation of the 11 year sunspot cycle. This is true regarding the magnitude of the
322    individual cycles as well as their length or repetition rate. Only 11 of the 28 sunspot cycles since
323    1700 A.D. have $^{10}$Be flux maxima at times when the sunspot number is a minimum as would be
324    expected from modulation theory.

325    When the first detailed $^{10}$Be measurements from polar ice cores were reported (e.g., Beer,
326    et al., 1990) there was the hope that this ice core data could provide a "monitor" of past solar
327    activity as it effects cosmic ray intensities incident on the Earth, in much the same way as
328    neutron monitors are used to monitor this solar activity in the modern era (Beer, 2000). This
329    "concept" with its 1:1 correspondence between $^{10}$Be production and $^{10}$Be in ice cores, has since
330    been used extensively to interpret historical $^{10}$Be ice core data in terms of changes in heliospheric
331    conditions and their effect on cosmic ray intensities incident on the Earth. Our results show that,
332    given our current understanding (or lack of it) of the correspondence between $^{10}$Be production,
333    sunspot numbers and the $^{10}$Be observed in ice cores, this is really not a reliable "concept" to use
334    for historical extrapolation. The sunspot number itself remains the best indicator of cyclic (11
335    year) solar activity after ~1700 A.D.

336    The level of $^{10}$Be production in the Earth's atmosphere at the minimum of the 5 most
337    recent solar activity cycles is the same to within ±5%. If this production level has changed in the
338    past, then the present uncertainties in the $^{10}$Be ice core data are such that the magnitude of these
339    changes cannot be robustly determined. Calculations also show that if the LIS remains the same
340    as it is now, the $^{10}$Be production should never increase by a factor of more than 1.3-1.5 over the
341    average modern values at sunspot minimum at any time in the recent past, whereas increases by
342    a factor ~2.0 have been observed in the $^{10}$Be ice core data.

343    In order that ice core data may in the future be utilized to provide an accurate historical
344    record of cosmic ray intensity incident on the Earth's polar atmosphere, new measurements or
345    analyses need to be made. The measurements would need to include comprehensive yearly

346 measurements of $^{10}$Be covering the last 60 years (several 11 year cycles) up to and including

347 2009, in several nearby (within a few km) ice cores in order to isolate and understand the origin

348 of the large fluctuations observed in the individual yearly ice core data. This would be an

349 extension of the earlier measurements of Moraal, et al., 2005.

350

# References

Bard, E., G. Raisbeck, F. Yiou and J. Jouzel, (1997), Solar modulation of cosmogenic nuclide production over the last millennium: Comparison between [14]C and [10]Be records, Earth Planet, Sci. Lett., 150, 453-462

Beer, J., et al., (1990), Use of [10]Be in polar ice to trace the 11-year cycle of solar activity, Nature, 347, 164-166

Beer, J., S. Tobias and N. Weiss, (1998), An active sun through the Maunder minimum, Solar Phys., 181, 237-249

Beer, J., (2000), Neutron monitor records in broader historical context, Space Science Reviews, 93, 89-100

Berggren, A.M., et al., (2009), A 600-year annual [10]Be record from the NGRIP ice core, Greenland, Geophys. Res. Lett., 36, L11801, doi:10.1029/2009GL038004

Field, C.V., G.A. Schmidt and D.T. Shindell, (2009), Interpreting [10]Be changes during the Maunder Minimum, J. Geophys Res., 114, D02113, doi:10.1029/2008JD010578

Lal, D., (1988), Theoretically expected variations in the terrestrial cosmic-ray production rates of isotopes, Proceedings of the International School of Physics "Enrico Fermi", Course XCV, Edited by G. Cini Castagnoli, p.216, North Holland Publishing

McCracken, K.G., F.B. McDonald, J. Beer, G. Raisbeck and F. Yiou, (2004), A phenomenological study of the long term cosmic ray modulation, 850-1958 AD, J. Geophys. Res., 109, A12103, doi:10.1029/2004JA010685

Moraal, H., et al., (2005), [10]Be concentration in the Ice Shelf of Queen Maude Land, Antarctica, South African Journal of Science, 101, 299-301

Nikitin, J., J. Stozkov, V. Okhlopkov and N. Svinzhevsky, (2005), Do Be-10 and C-14 give us information about cosmic rays in the past?, Proc., 20[th] ICRC, Pune, 2, 243-246

Pedro, J., et al., (2006), Evidence for climatic modulation of the [10]Be solar activity proxy, J. Geophys. Res., 111, D21103, doi:10.1029/2005JF006764

Steinhilber, F., J.A. Abreu, J. Beer and K.G. McCracken, (2010) Interplanetary magnetic field during the past 9300 years inferred from cosmogenic radionuclides, J. Geophys. Res., 115, A01104, doi:10.1029/2009JA014193

380  Stozhkov, Yu, I., (2007), What can be extracted from data on the concentrations of Be-10 and C-
381      14 natural radionuclides?, Bulletin of the Lebedev Physics Institute, <u>34</u>, #5, 135-141 at
382      Allerton Press Inc.

383  Usoskin, I.G., K. Horiuchi, S. Solanki, G.A. Kovaltsov and E. Bard, (2009), On the common
384      solar signal in different cosmogenic isotope data sets, J. Geophys. Res., <u>114</u>, A0.112, doi:
385      10.1029/2008JA013888

386  Webber, W.R., P.R. Higbie and K.G. McCracken, (2007), The production of the cosmogenic
387      isotopes $^3$H, $^7$Be, $^{10}$Be and $^{36}$Cl in the Earths atmosphere by solar and galactic cosmic rays, J.
388      Geophys. Res., <u>112</u>, A10106, doi:10.1029/2007JA012499

389  Webber, W.R. and P.R. Higbie, (2010b) What Voyager cosmic ray data in the outer heliosphere
390      tells us about $^{10}$Be production in the Earths polar atmosphere in the recent past, J. Geophys
391      Res., in press

392  Webber, W.R. and P.R. Higbie, (2010a) A comparison of new calculations of $^{10}$Be production in
393      the Earths polar atmosphere by cosmic rays with $^{10}$Be concentration measurements in polar
394      ice cores between 1939–2005 – A troubling lack of concordance: Paper #1,
395      http://arxiv.org/abs/1003.4989

396

| | | | | | | | | |
|---|---|---|---|---|---|---|---|---|
| **TABLE 1** | | | | | | | | |
| YEAR | NGRIP CONC x 10^4 | NGRIP FLUX x10⁻³ | WH POLAR PRODUC | DYE-3 CONC x 10⁴ | DYE-3 FLUX x 10⁻³ | (B+2E) 2 | SUNSPOT# NOAA | NM WRW 1954=100 |
| 1939.5 | 0.93 | 5.62 | 4.18 | 0.51 | 11.7 | 0.98 | 89 | |
| 1940.5 | 1.31 | 9.10 | 4.45 | 0.56 | 11.6 | 1.215 | 68 | |
| 1941.5 | 1.19 | 6.00 | 4.40 | 0.64 | 14.1 | 1.285 | 47 | |
| 1942.5 | 1.01 | 5.60 | 4.62 | 0.89 | 15.6 | 1.43 | 31 | |
| 1943.5 | 1.61 | 7.60 | 4.70 | 1.11 | 21.0 | 1.915 | 16 | |
| 1944.5 | 1.63 | 7.70 | 5.22 | 0.57 | 7.8 | 1.385 | 9 | |
| 1945.5 | 1.33 | 7.35 | 5.04 | 0.61 | 12.0 | 1.34 | 33 | |
| 1946.5 | 1.12 | 7.75 | 3.32 | 0.65 | 17.3 | 1.245 | 91 | |
| 1947.5 | 1.10 | 5.45 | 2.75 | 0.57 | 13.8 | 1.14 | 149 | |
| 1948.5 | 1.16 | 7.90 | 3.18 | 0.905 | 12.6 | 1.485 | 137 | |
| 1949.5 | 1.48 | 7.90 | 4.25 | 0.61 | 8.8 | 1.35 | 136 | |
| 1950.5 | 1.02 | 4.52 | 4.60 | 1.18 | 17.8 | 1.705 | 84 | |
| 1951.5 | 2.44 | 14.00 | 4.30 | 1.02 | 15.6 | 2.24 | 69.5 | 0.949 |
| 1952.5 | 1.50 | 8.00 | 4.56 | 1.02 | 14.2 | 1.77 | 31.5 | 0.963 |
| 1953.5 | 1.46 | 7.10 | 4.63 | 1.24 | 19.6 | 1.985 | 14 | 0.978 |
| 1954.5 | 1.26 | 6.80 | 5.05 | 0.77 | 14.7 | 1.445 | 4.4 | 1.00 |
| 1955.5 | 1.36 | 12.30 | 4.90 | 1.01 | 15.8 | 1.71 | 38 | 0.996 |
| 1956.5 | 1.42 | 6.80 | 4.23 | 0.57 | 10.5 | 1.28 | 142 | 0.951 |
| 1957.5 | 1.60 | 7.40 | 2.83 | 0.895 | 12.6 | 1.685 | 190 | 0.837 |
| 1958.5 | 1.00 | 5.10 | 2.70 | 0.41 | 6.0 | 0.925 | 185 | 0.826 |
| 1959.5 | 1.23 | 5.85 | 2.83 | 0.49 | 10.2 | 1.125 | 159 | 0.832 |
| 1960.5 | 1.25 | 5.90 | 2.92 | 0.41 | 9.5 | 1.05 | 112 | 0.845 |
| 1961.5 | 1.05 | 4.55 | 3.52 | 0.40 | 7.5 | 0.94 | 54 | 0.892 |
| 1962.5 | 1.66 | 9.10 | 3.90 | 0.50 | 7.8 | 1.28 | 37.5 | 0.914 |
| 1963.5 | 2.08 | 10.20 | 4.26 | 0.62 | 7.6 | 1.66 | 28 | 0.945 |
| 1964.5 | 1.45 | 7.90 | 4.74 | 0.69 | 15.5 | 1.235 | 10.2 | 0.978 |
| 1965.5 | 3.20 | 14.20 | 5.02 | 1.03 | 11.8 | 2.63 | 15 | 0.998 |
| 1966.5 | 1.74 | 8.40 | 4.54 | 0.593 | 8.2 | 1.478 | 47 | 0.962 |
| 1967.5 | 1.69 | 7.40 | 4.02 | 0.83 | 12.6 | 1.695 | 94 | 0.926 |
| 1968.5 | 1.44 | 8.60 | 3.62 | 0.47 | 5.8 | 1.175 | 106 | 0.896 |
| 1969.5 | 1.14 | 5.60 | 3.40 | 0.50 | 7.2 | 1.115 | 105 | 0.883 |
| 1970.5 | 1.04 | 5.20 | 3.48 | 0.40 | 9.8 | 0.92 | 104 | 0.889 |
| 1971.5 | 1.35 | 8.35 | 4.40 | 0.50 | 7.2 | 1.19 | 67 | 0.950 |
| 1972.5 | 1.76 | 8.10 | 4.65 | 0.35 | 9.6 | 1.23 | 69 | 0.971 |
| 1973.5 | 1.61 | 9.85 | 4.68 | 0.695 | 11.8 | 1.485 | 38 | 0.971 |
| 1974.5 | 1.47 | 6.05 | 4.50 | 1.00 | 14.6 | 1.755 | 34.5 | 0.961 |
| 1975.5 | 1.31 | 8.90 | 4.90 | 0.77 | 11.7 | 1.465 | 15.5 | 0.983 |
| 1976.5 | 1.38 | 9.60 | 4.95 | 0.59 | 9.1 | 1.31 | 12.5 | 0.986 |
| 1977.5 | 1.84 | 8.10 | 4.83 | 0.57 | 10.3 | 1.49 | 27.5 | 0.985 |
| 1978.5 | 1.39 | 7.10 | 4.40 | 0.62 | 11.8 | 1.345 | 92.5 | 0.955 |
| 1979.5 | 1.56 | 9.10 | 3.72 | 0.61 | 8.1 | 1.39 | 155 | 0.902 |
| 1980.5 | 0.96 | 4.70 | 3.39 | 0.60 | 8.3 | 1.095 | 154 | 0.879 |
| 1981.5 | 1.28 | 6.80 | 3.04 | 0.65 | 11.8 | 1.345 | 141 | 0.854 |
| 1982.5 | 1.47 | 7.20 | 2.93 | 0.66 | 13.4 | 1.395 | 116 | 0.848 |
| 1983.5 | 1.71 | 5.35 | 3.35 | 1.01 | 14.4 | 1.845 | 68.5 | 0.879 |
| 1984.5 | 1.52 | 10.10 | 3.59 | 0.50 | 6.0 | 1.26 | 46 | 0.889 |
| 1985.5 | 2.18 | 8.10 | 4.22 | 0.48 | 4.6 | 1.55 | 17 | 0.935 |
| 1986.5 | 1.46 | 8.75 | 4.92 | | | | 13.5 | 0.980 |
| 1987.5 | 1.11 | 7.00 | 4.96 | | | | 29.5 | 0.983 |
| 1988.5 | 1.65 | 6.10 | 3.93 | | | | 100 | 0.913 |
| 1989.5 | 1.41 | 6.05 | 2.69 | | | | 158 | 0.816 |
| 1990.5 | 0.86 | 6.80 | 2.63 | | | | 142 | 0.806 |
| 1991.5 | 1.11 | 6.60 | 2.61 | | | | 146 | 0.809 |
| 1992.5 | 1.52 | 7.90 | 3.62 | | | | 94.5 | 0.890 |
| 1993.5 | 1.49 | 7.70 | 4.18 | | | | 54.5 | 0.939 |
| 1994.5 | 1.59 | 9.10 | 4.34 | | | | 30 | 0.953 |
| 1995.5 | | | 4.66 | | | | 17.5 | 0.980 |
| 1996.5 | | | 4.90 | | | | 8.6 | 0.989 |
| 1997.5 | | | 4.94 | | | | 21.4 | 0.993 |
| 1998.5 | | | 4.63 | | | | 64 | 0.971 |
| 1999.5 | | | 4.08 | | | | 93 | 0.934 |
| 2000.5 | | | 3.08 | | | | 120 | 0.866 |
| 2001.5 | | | 3.23 | | | | 111 | 0.873 |
| 2002.5 | | | 3.20 | | | | 104 | 0.871 |
| 2003.5 | | | 3.03 | | | | 63 | 0.861 |
| 2004.5 | | | 4.00 | | | | 40.5 | 0.922 |
| 2005.5 | | | 4.12 | | | | 30 | 0.928 |
| 2006.5 | | | 4.70 | | | | 15.5 | 0.975 |
| 2007.5 | | | 4.92 | | | | 6.5 | 0.991 |
| 2008.5 | | | 5.03 | | | | 2 | 0.998 |
| 2009.5 | | | 5.30 | | | | 0.5 | 1.016 |
| 2010.5 | | | | | | | | |

398  Conc. Units         = atoms · g$^{-1}$

399  Flux units          = atoms · cm$^{2}$s$^{-1}$

400  Polar prod units    = atoms · cm$^{-2}$s$^{-1}$

401

**Figure Captions**

**Figure 1:**  Yearly average normalized high latitude NM rates (1954=100.0), yearly average $^{10}$Be production (on right hand axis).

**Figure 2A:**  Scatter plot – normalized NM rate vs. $^{10}$Be production.

**Figure 2B:**  Scatter plot – sunspot number vs. $^{10}$Be production.

**Figure 3A:**  Scatter plot – NGRIP flux vs. $^{10}$Be production (normalized slope of regression line=0.503).

**Figure 3B:**  Scatter plot – Dye-3 concentration vs. $^{10}$Be production (normalized slope of regression line=0.573).

**Figure 3C:**  Scatter plot – Dye-3 flux vs. $^{10}$Be production (normalized slope of regression line=0.437).

**Figure 4A:**  Scatter plot – Average NGRIP plus Dye-3 concentration vs. $^{10}$Be production (normalized slope of regression line=0.544).

**Figure 4B:**  Scatter plot – Dye-3 flux vs. NGRIP flux (normalized slope of regression line=0.003).

**Figure 5:**  Five year running average NGRIP flux data from Berggren, et al., 2009, from 1600 A.D. to the present.  Also shown is the yearly average sunspot number (on an inverted scale) from http://www.ngdc.noaa.gov/stp/SOLAR/ftpsunspotnumber.html#american.  The shaded region shows the maximum limits for $^{10}$Be flux values for a full LIS spectrum incident on the Earths atmosphere for different assumptions regarding the correspondence between $^{10}$Be production and $^{10}$Be ice core measurements.  The dashed lines for the time period after 1940 represent the average maximum and minimum values of $^{10}$Be flux for the 5 most recent solar cycles.  See text for a description of the blue lines.

**Figure 6:**  Yearly average values for cosmic ray production of $^{10}$Be (from Webber, Higbie and McCracken, 2007) in blue; Sunspot number (from http://www.ngdc.noaa.gov/stp/SOLAR/ftpsunspotnumber.html#american) on an inverted scale (in red) and five year running average $^{10}$Be flux from the NGRIP ice core (Berggren, et al, 2009) (in black).  Data is shown from ~1930 to the present.

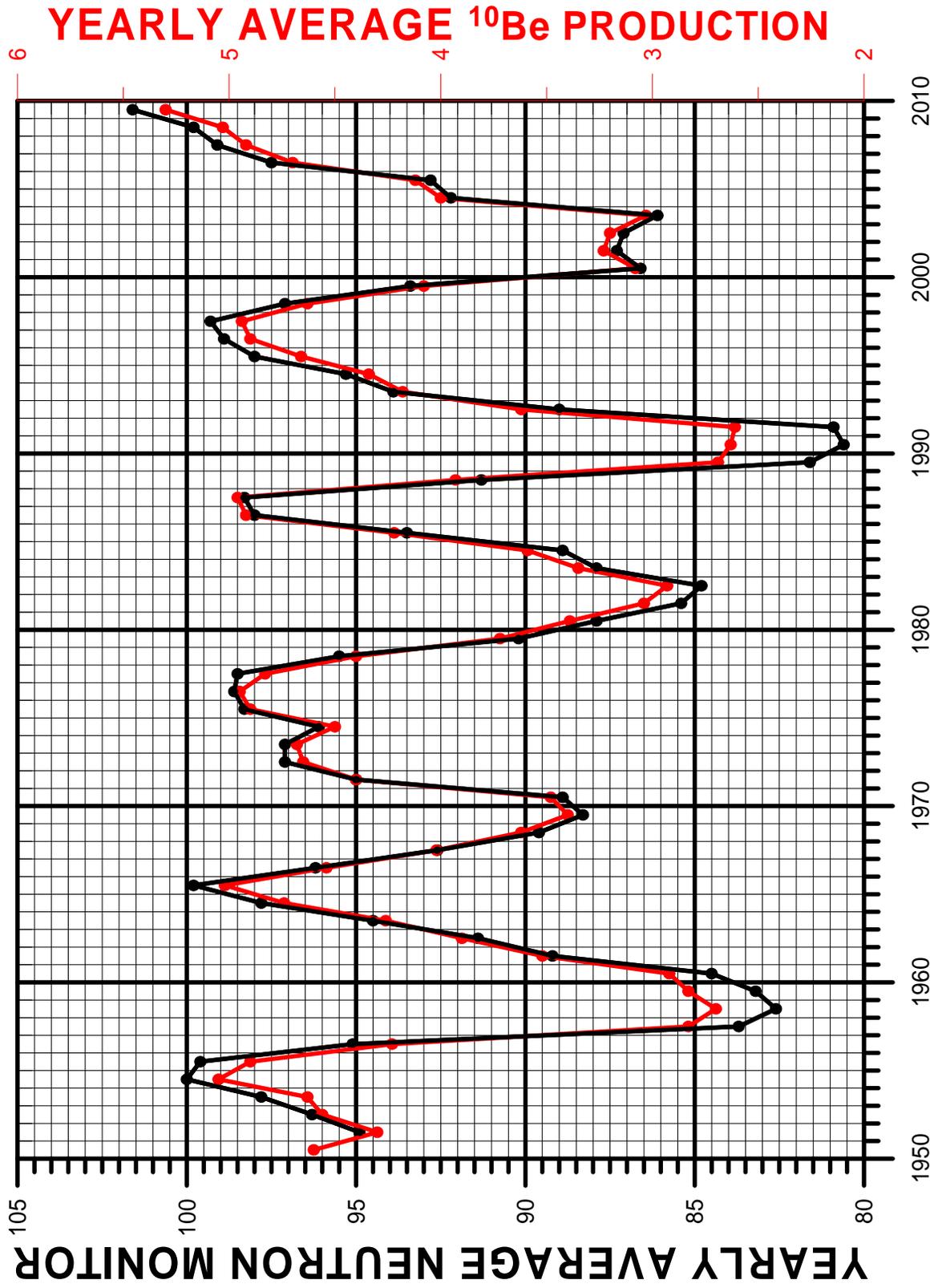

**YEARLY AVERAGE $^{10}$Be PRODUCTION**

**TIME**

**FIGURE 1**

**YEARLY AVERAGE NEUTRON MONITOR**

431

432

433
434

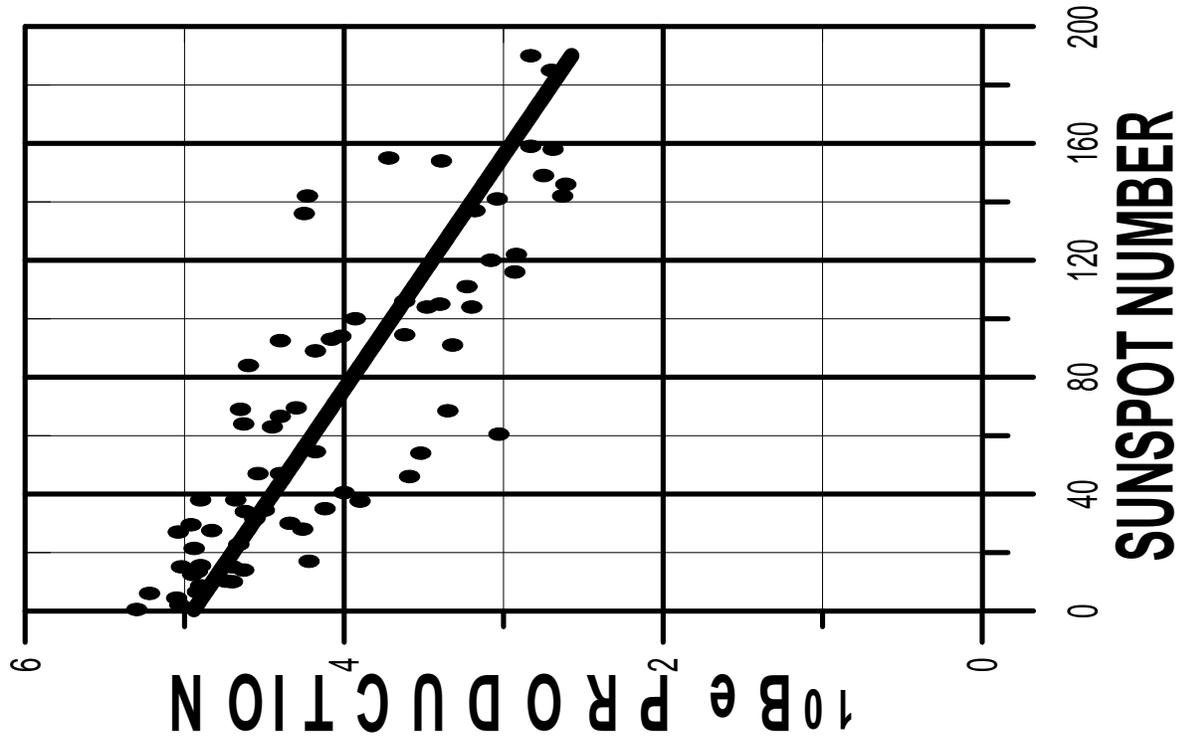

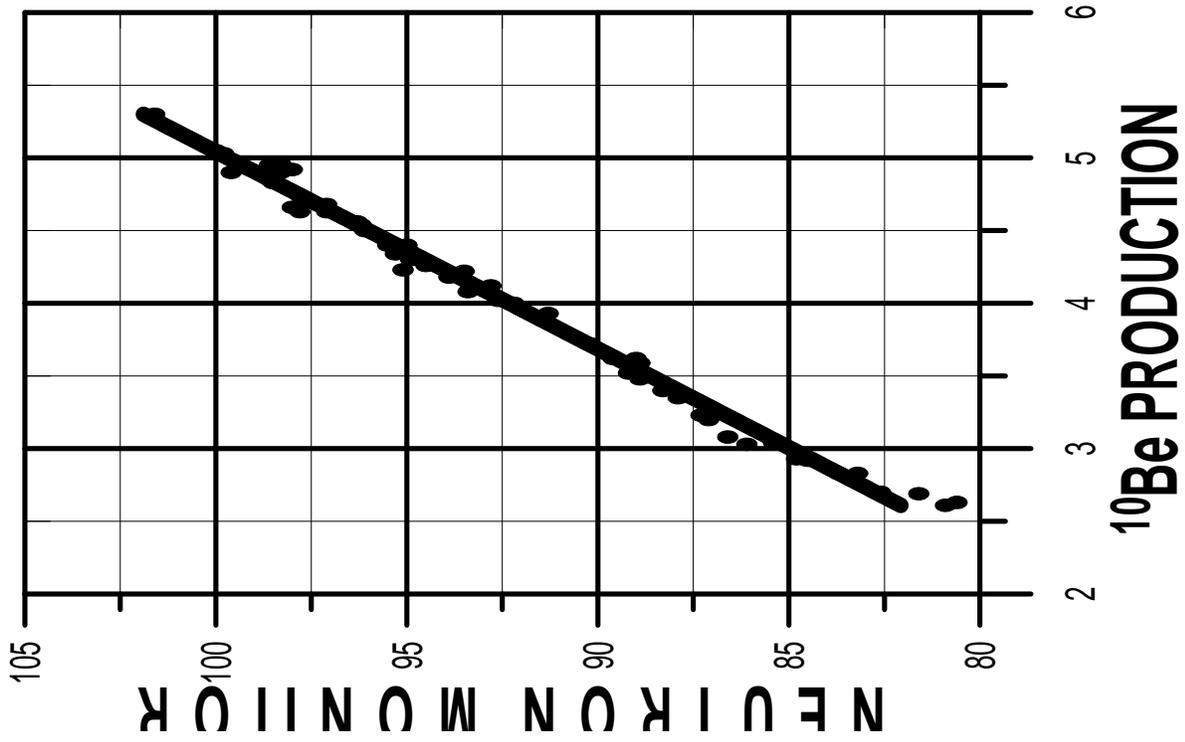

435
436

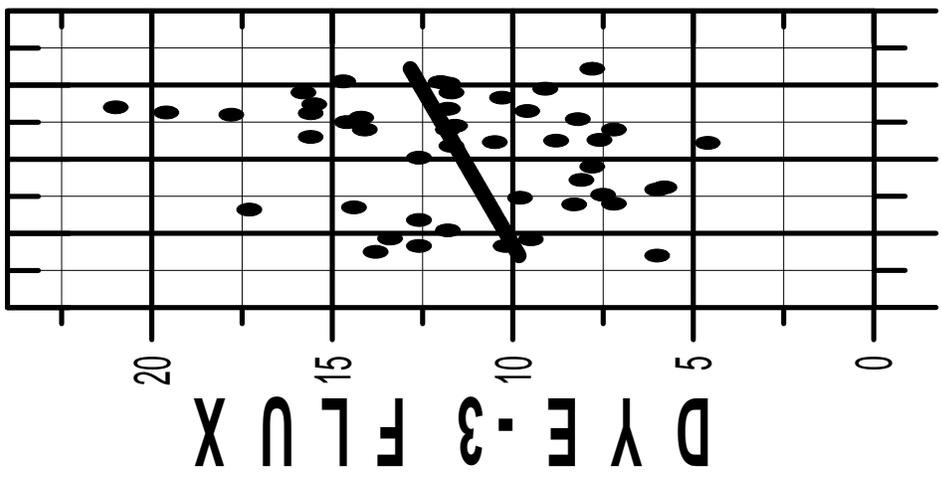

NO₃⁻ FLUX

$^{10}$Be PRODUCTION

FIGURE 3A

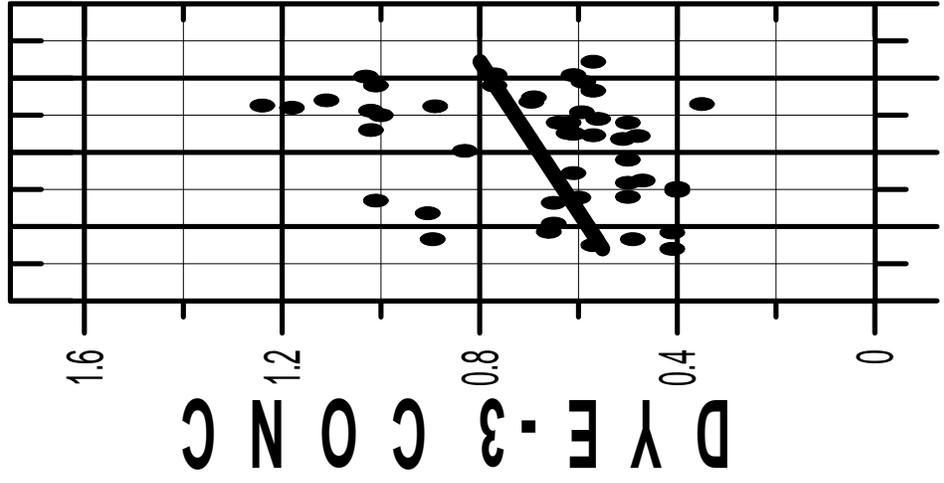

DYE-3 CONC

$^{10}$Be PRODUCTION

FIGURE 3B

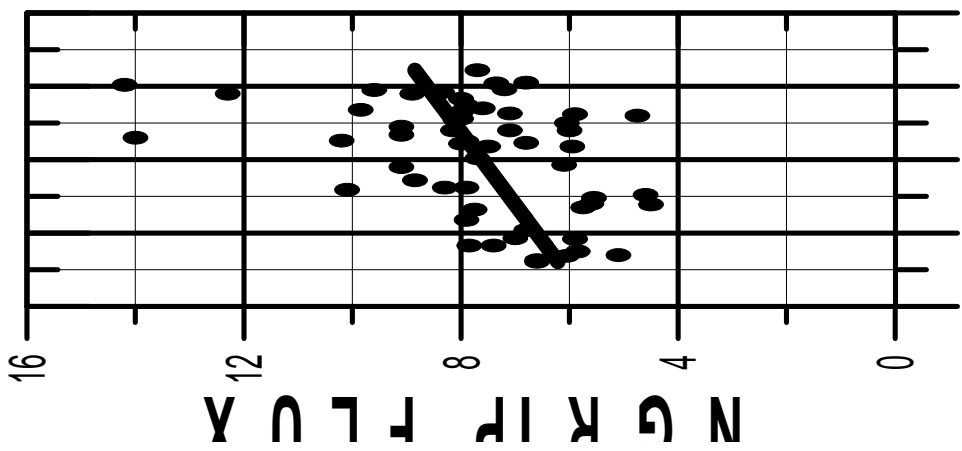

DYE-3 FLUX

$^{10}$Be PRODUCTION

FIGURE 3C

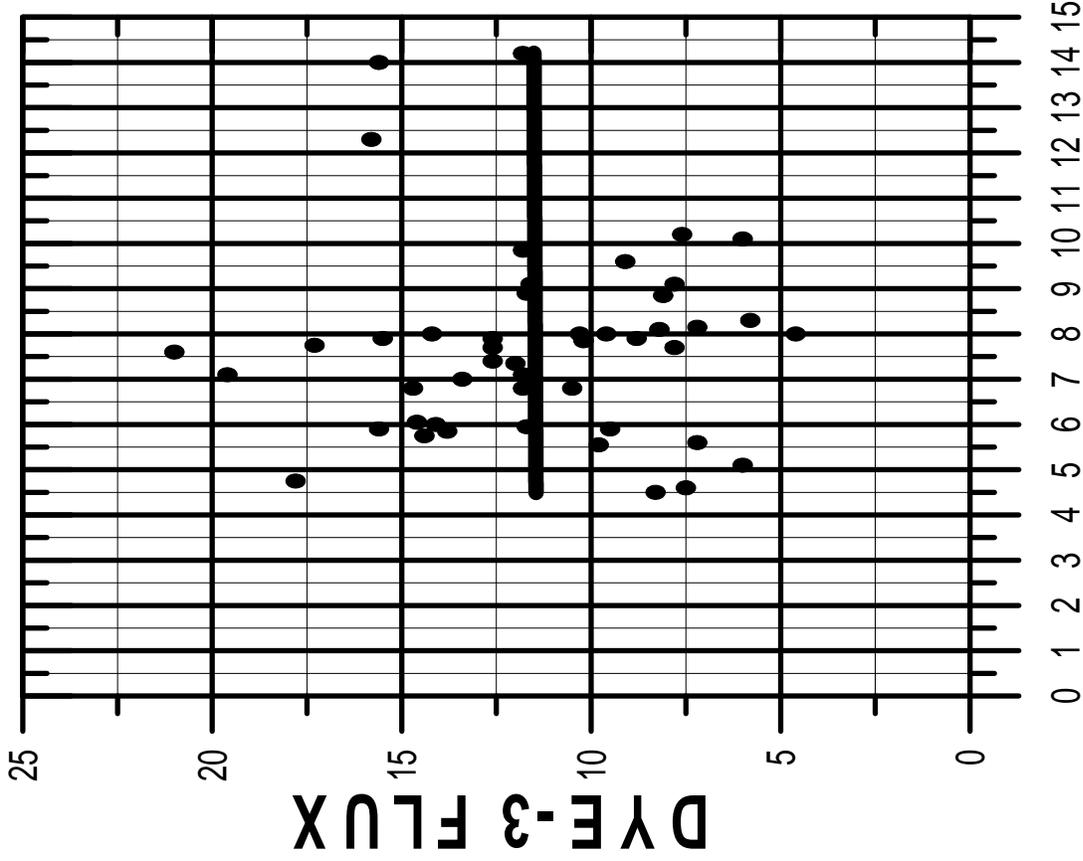

**NGRIP FLUX**

**FIGURE 4B**

**DYE-3 FLUX**

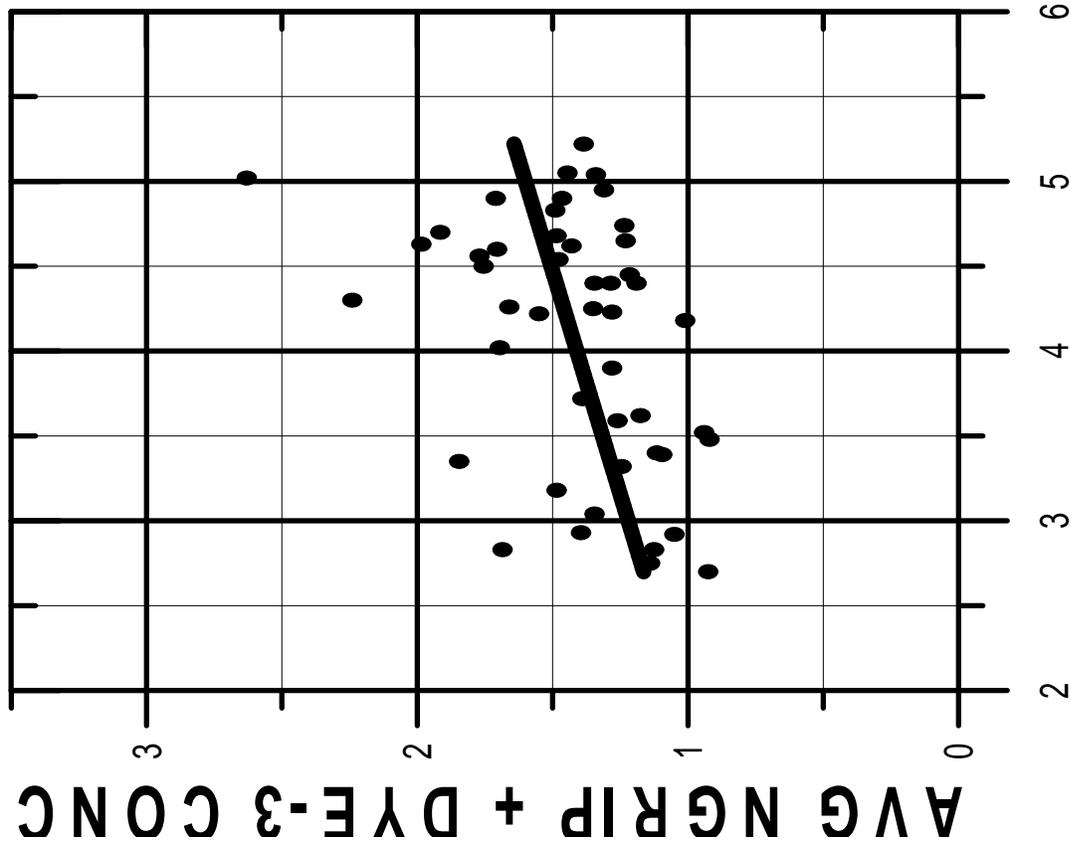

**¹⁰Be PRODUCTION**

**FIGURE 4A**

**AVG NGRIP + DYE-3 CONC**

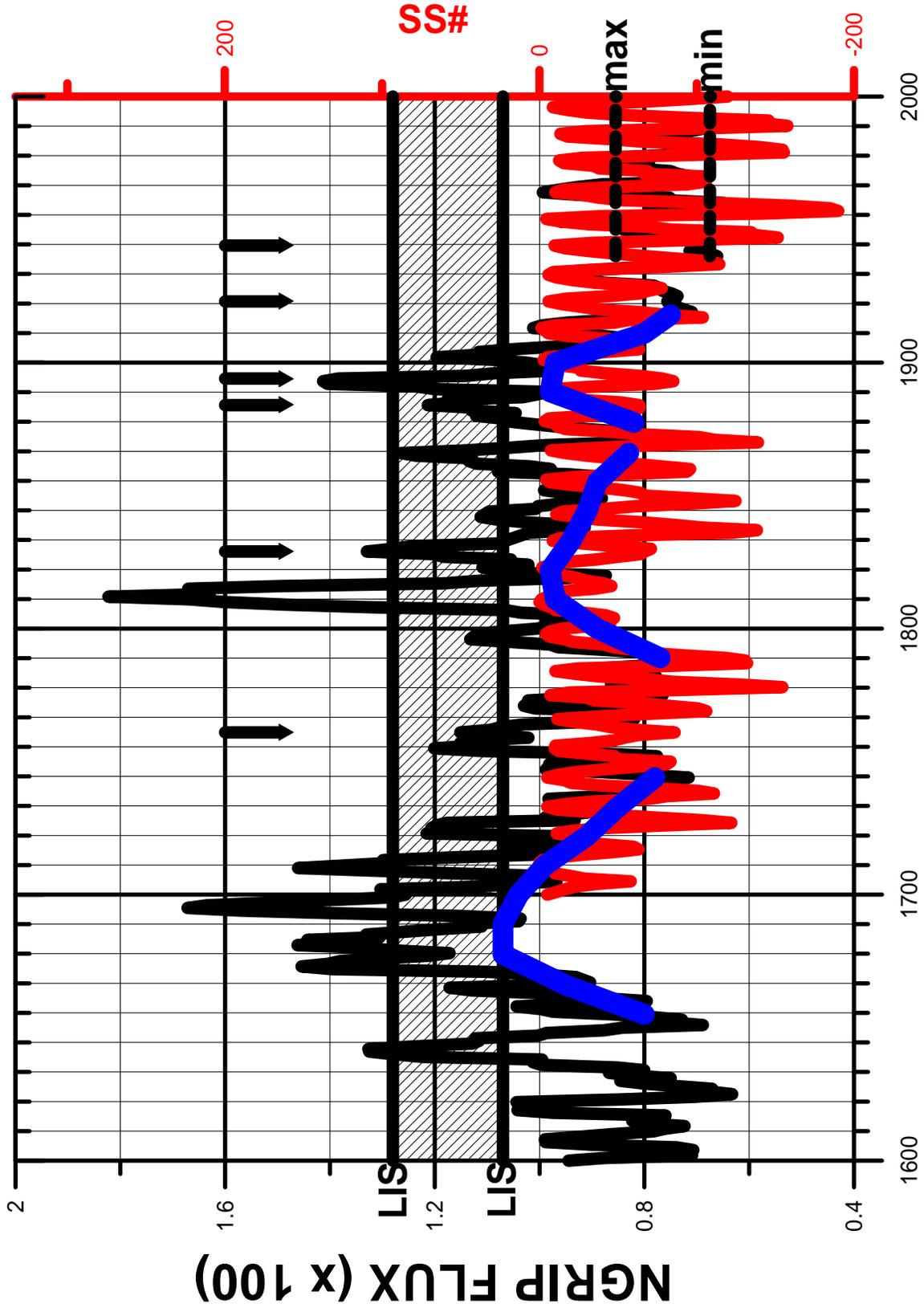

FIGURE 5

438

439

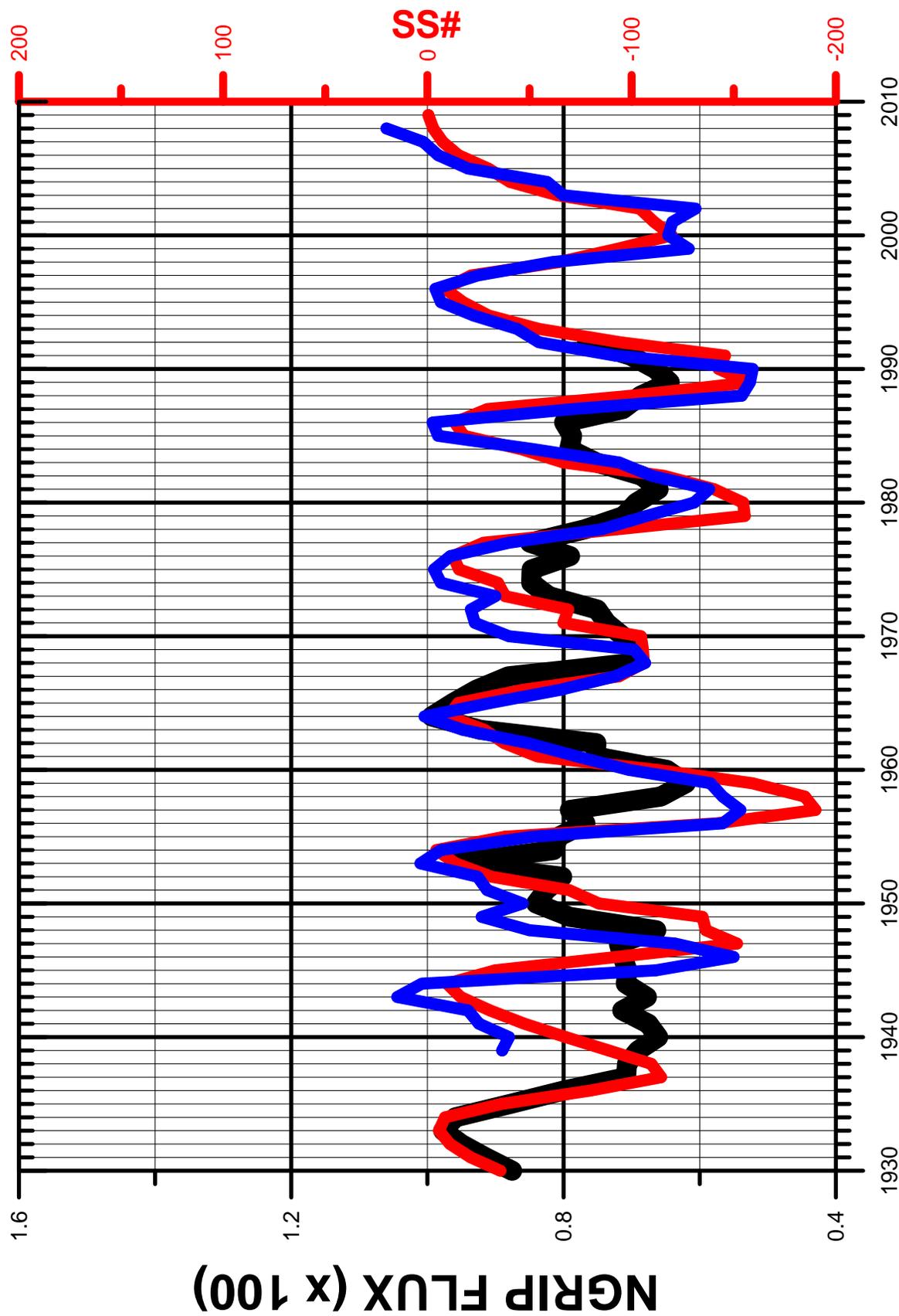

FIGURE 6